\documentclass[journal]{IEEEtran}

\ifCLASSINFOpdf
\else
\fi

\usepackage{amsmath,mathrsfs,amsfonts,amssymb,amsthm}
\usepackage{color,cite}
\usepackage{subfigure}
\usepackage{tabularx}
\usepackage{graphics,graphicx}

\begin{document}
\title{Volterra Series Based Time-domain Macro-modeling of Nonlinear Circuits}

\author{Xiaoyan~Y.~Z.~Xiong,~\IEEEmembership{Member,~IEEE,}
        Li~Jun~Jiang,~\IEEEmembership{Senior Member,~IEEE,}
        Jos\'{e}~E.~Schutt-Ain\'{e},~\IEEEmembership{Fellow,~IEEE,}
        and Weng~Cho~Chew~\IEEEmembership{Fellow,~IEEE,}
\thanks{X.~Y.Z.~Xiong and L.~J.~Jiang  are with the Department of Electrical and Electronic Engineering, University of Hong Kong, Pokfulam Road, Hong Kong, China (e-mail: xyxiong@eee.hku.hk; jianglj@hku.hk).}
\thanks{J. Schutt-Ain\'{e} and W. C. Chew are with the Department of Electrical and Computer Engineering,
University of Illinois at Urbana-Champaign, Urbana, IL 61801 USA (e-mail: jschutt@emlab.uiuc.edu; w-chew@uiuc.edu). W. C. Chew was with the Faculty of Engineering, University of Hong Kong, Hong Kong.}
}


\maketitle

\begin{abstract}
Volterra series representation is a powerful mathematical model for nonlinear circuits.  However, the difficulties in determining higher-order Volterra kernels limited its broader applications.  In this work, a systematic approach that enables a convenient extraction of Volterra kernels from X-parameters is presented. A concise and general representation of the output response due to arbitrary number of input tones is given. The relationship between Volterra kernels and X-parameters is explicitly formulated. An efficient frequency sweep scheme and an output frequency indexing scheme are provide. The least square linear regression method is employed to separate different orders of Volterra kernels at the same frequency, which leads to the obtained Volterra kernels complete. The proposed Volterra series representation based on X-parameters is further validated for time domain verification. The proposed method is systematic and general-purpose. It paves the way for time domain simulation with X-parameters and constitutes a powerful supplement to existing blackbox macro-modeling methods for nonlinear circuits.
\end{abstract}

\begin{IEEEkeywords}
Volterra series, X-parameters, nonlinear circuits, blackbox macro-modeling.
\end{IEEEkeywords}

%

\section{Introduction}\label{Sec1}
\IEEEPARstart{T}{he} macro-modeling of nonlinear devices and systems is a topic of growing interest due to the dramatic increases in the complexity and size of modern systems~\cite{ref01,ref02_1,ref02_2,ref02_3,ref03,ref04}. The basic idea of macro-modeling of a circuit system is to replace the original system by an approximating system which requires much less design time and fewer resources. Building macro-models is key for enabling complete system verification and high-level design exploration.
Recently developed nonlinear macro-modeling methods can be largely categorized into two major classes: nonlinear model order reduction (MOR) and nonlinear blackbox macro-model generation. Both approaches are substantially harder than their linear counterparts. The first type works directly on the SPICE level schematics, i.e., state equation derived from modified nodal analysis (MNA) method. For nonlinear MOR, a traditional method is to first linearize the system then extend projection-based MOR techniques for linear systems to accommodate nonlinear systems~\cite{ref05,ref06_1,ref07,ref08_1}. By doing this, the nonlinear MOR task is re-cast as the reduction of a series of linear systems. However, this method suffers from the exponentially increased dimension. Its {relevant system transfer function is  prohibitively too difficult to expand and to generate moments, essentially limiting the approach to weakly nonlinear systems. The trajectory-based piecewise-linear (TPWL) approximation approach models a nonlinear system using a collection of weighted linear models based upon a state trajectory generated by a training input~\cite{ref09}. Then each linear model is reduced using linear MOR techniques. This approach has the potential capability to handle large nonlinearities, but is limited by the training input dependency.

The above-mentioned nonlinear MOR methods are based on SPICE level modeling. However, in some situations, it is difficult to obtain the SPICE model due to intellectual property restrictions and limited information. Some systems cannot be described by SPICE level models because of coupling effects, distributed elements, and higher-order modes (excited by via, connector), etc.~\cite{ref11_1}. Consequently, blackbox macro-modeling becomes a viable alternative. The goal of blackbox macro-modeling is to find a mathematical relation that can reproduce the electrical behavior at the ports without any assumption about the device's internal structure. However, it is difficult to find accurate and efficient models to characterize the nonlinear behavior of devices under arbitrary loads and input signals. S-parameters are the network parameters that have been used widely as a blackbox macromodel in the signal integrity and RF/microwave frequency domains~\cite{ref12,ref13,ref14_1,ref14_2}. But the applicability of S-Parameters has been limited to small signals and linear behaviors.

The recently developed X-parameters from the poly-harmonic distortion (PHD) model~\cite{ref15} are a superset of S-parameters. They describe the relationships between incident and scattered waves by using not only port-to-port but also harmonic-to-harmonic interactions under certain large signal operating points (LSOPs). They have been successfully used to describe various nonlinear devices~\cite{ref16}. In the descriptive function concept, input signals are restricted to fundamental components consisting of LSOPs superposed with small harmonics. Consequently, X-parameters, essentially a frequency-domain tool, cannot support time domain simulations and have difficulty handling input signals with high peak-to-average ratios that will excite the device over broad linear and nonlinear operating ranges.

Dynamic X-parameters~\cite{ref17_1,ref17_2,ref17_3} is a fundamental extension of X-parameters to modulation-induced baseband memory effects. It can be used to model hard nonlinear behavior and long term memory effects and is valid for all possible modulation formats, for all possible peak-to-average ratios and for a wide range of modulation bandwidths. The model can be implemented in a commercial complex envelope simulator. The core of dynamic X-parameters is the memory kernel function derived from hidden variables concept. The memory kernel can be regarded as the nonlinear impulse response of the system and can be uniquely identified from the set of complex envelope time domain measurement from initial states  to final states. However, the current dynamic X-parameters model is defined in the envelope domain under several basic assumptions. The incident waves  and scattered waves are restricted to be complex envelope representations of modulated carriers.

The Volterra series representation is another popular black-box macro-modeling approach for describing nonlinear devices with memory~\cite{ref18,ref19,ref20}. It can support time domain simulation with arbitrary input and is valid for signals that can excite both linear and nonlinear responses. Without knowing the state equation, the difficulty in determining higher-order Volterra kernels has restricted its application.
We previously proposed a method to get Volterra kernels from X-parameters~\cite{ref17}. However, in~\cite{ref17}, the Volterra kernel model is used as a frequency domain model. The input is restricted to harmonic input. It provides no more information than X-parameters and is unable to conduct time domain simulation. In this paper, we have extended the Volterra kernel model to time domain simulation. To make the paper more easily readable and self-contained, some formulas are rewritten and additional formulas are provided with detailed explanation. The accuracy and capability of Volterra kernel model for time domain simulation have been verified. A concise and general representation of the output responses due to arbitrary number of input tones is given. The generalized relationship between Volterra kernels and X-parameters is explicitly formulated. In addition, the requirements of the input signal are discussed in detail. An efficient frequency sweep scheme and an output frequency indexing scheme are provide. With these schemes and the symmetry property of Volterra kernels, there is no need to determine the irreducible frequency sweeping region, which makes the kernel determination procedure more convenient for computer programming. The Volterra kernels are extracted only once, which can be repeatedly used for different types of input. The proposed method is systematic and general-purpose. It is very useful for the blackbox macro-modeling of nonlinear circuits.

The organization of the paper is as follows. Section II provides a brief description of  Volterra series theory. Section III gives the X-parameters formalism with incommensurate multi-tone input. Section IV presents the detailed technical description of the Volterra kernel extraction process. In Section V, numerical examples of extracting Volterra series from the X-parameters are provided. Time domain outputs are presented to validate the proposed method. Finally, a conclusion is given in Section VI.

\section{Volterra Series}\label{Sec2}
Volterra series have been widely used to characterize nonlinear systems with memory~\cite{ref22}. For a system with input $u(t)$, the output $y(t)$ can be expressed using the expansion
\begin{eqnarray}\label{eq1}
y(t)=\sum_{n=1}^{\infty}y_n(t)
\end{eqnarray}
with
\begin{eqnarray}\label{eq2}
y_n(t)=\frac{1}{n!}\int_{-\infty}^{+\infty}\ldots\int_{-\infty}^{+\infty}h_n(\tau_1,\ldots,\tau_n)\cdot\\ \nonumber
u(t-\tau_1)\ldots u(t-\tau_n)d\tau_1\ldots d\tau_n \nonumber
\end{eqnarray}
where $h_n(\tau_1,\ldots,\tau_n)$ is the $n$th-order time domain Volterra kernel or impulse response. In particular, $y_1(t)$ is the usual first-order convolution having its frequency domain representation
\begin{eqnarray}\label{eq3}
Y_1(\omega)=H_1(\omega)U(\omega)
\end{eqnarray}
where $H_1(\omega)=\int_{-\infty}^{+\infty}h_1(\tau)e^{-j\omega\tau}d\tau$ is the linear transfer function or the first-order Volterra kernel. $U(\omega)$ is the Fourier transform of $u(t)$. However, the nonlinear higher-order output cannot be written in a form similar to \eqref{eq3}. By replacing the single time axis by multiple time axes, \eqref{eq2} becomes
\begin{eqnarray}\label{eq4}
y_n(t_1,\ldots,t_n)=\frac{1}{n!}\int_{-\infty}^{+\infty}\ldots\int_{-\infty}^{+\infty}h_n(\tau_1,\ldots,\tau_n)\cdot\\ \nonumber
u(t_1-\tau_1)\ldots u(t_n-\tau_n)d\tau_1\ldots d\tau_n \nonumber
\end{eqnarray}
The frequency domain representation of \eqref{eq4} can be conveniently written in a form similar to \eqref{eq3}
\begin{eqnarray}\label{eq5}
Y_n(\omega_1,\ldots,\omega_n)=H_n(\omega_1,\ldots,\omega_n)U(\omega_1)\ldots U(\omega_n)
\end{eqnarray}
with the nonlinear transfer function $H_n$ defined as
\begin{eqnarray}\label{eq6}
H_n(\omega_1,\ldots, \omega_n)=\int_{-\infty}^{+\infty}\ldots\int_{-\infty}^{+\infty}h_n(\tau_1,\ldots,\tau_n)\cdot\\ \nonumber
e^{-j\omega_1\tau_1}\ldots e^{-j\omega_n\tau_n}d\tau_1\ldots d\tau_n\nonumber
\end{eqnarray}
To restore $y_n(t)$, one then evaluates along the diagonal line in the multi-time hyperplane
\begin{eqnarray}\label{eq7}
y_n(t)=y_n(t_1,\ldots,t_n)|_{t_1=t_2=\ldots=t_n=t}
\end{eqnarray}
where $y_n(t_1,\ldots,t_n)$ is the multidimensional inverse Fourier transform of $Y_n(\omega_1,\ldots,\omega_n)$.

\section{X-parameters Formalism}\label{Sec3}
X-parameters, derived from the polyharmonic distortion (PHD) model~\cite{ref15,ref23}, are a superset of S-parameters and
can be used to describe the behavior of nonlinear devices in the frequency domain. For better clarification and without losing generality, we take the incommensurate 3-tone excitation case to illustrate the formalism of X-parameters~\cite{ref24}.
Suppose the incident signal $A_q(t)$ at port $q$ has three large incommensurate fundamental
tones with frequencies $\omega_1,\,\omega_2\textrm{ and }\omega_3$. They are incommensurate if the ratio of their frequencies is irrational:
\begin{eqnarray}\label{eq8}
k_1\omega_1+k_2\omega_2+k_3\omega_3=0\Rightarrow k_1=k_2=k_3=0 \\ \nonumber
\textrm{for } k_1,k_2,k_3 \in \mathbb{Z}
\end{eqnarray}
The scattered signal at port $p$ contains numerous frequency components. They are the combinations of the input tones  $\omega=k_1\omega_1+k_2\omega_2+k_3\omega_3$ and can be indexed as $B_{p,[k_1,k_2,k_3]}$. Then, X-parameters are used to link the scattered signal $B$ with incident signal $A$ under linearization around specific LSOPs (amplitudes and phases of input tones, bias, loads, etc.).
\begin{eqnarray}\nonumber\label{eq9}
&&\!\!\!\!\!\!\!\!\!\!\!\!\!\!\!B_{p,[k_1,k_2,k_3]}=X_{p,[k_1,k_2,k_3]}^{(F)}P_{[1,0,0]}^{k_1} P_{[0,1,0]}^{k_2} P_{[0,0,1]}^{k_3}  \quad \quad \quad \quad \textrm{(9) } \\ \nonumber
&+&\!\!\!\!\sum_{q,k_1',k_2',k_3'} \Big[X_{p,[k_1,k_2,k_3];q,[k_1',k_2',k_3']}^{(S)}P_{[1,0,0]}^{k_1-k_1'} P_{[0,1,0]}^{k_2-k_2'} P_{[0,0,1]}^{k_3-k_3'} \\ \nonumber
&& \quad \quad \quad \quad   A_{q,[k_1',k_2',k_3']} \Big] \\ \nonumber
&+&\!\!\!\!\sum_{q,k_1',k_2',k_3'} \Big[X_{p,[k_1,k_2,k_3];q,[k_1',k_2',k_3']}^{(T)}P_{[1,0,0]}^{k_1+k_1'} P_{[0,1,0]}^{k_2+k_2'} P_{[0,0,1]}^{k_3+k_3'} \\ \nonumber
&& \quad \quad  \quad \quad  A_{q,[k_1',k_2',k_3']}^* \Big]  \nonumber
\end{eqnarray}
The $X^{(F)}$ term includes the information of the three large fundamental tones $A_{q,[1,0,0]}, A_{q,[0,1,0]}$  and $A_{q,[0,0,1]}$. $P_{[1,0,0]}$, $P_{[0,1,0]}$ and $P_{[0,0,1]}$ are the initial phases of input tones. $X^{(S)}$ and $X^{(T)}$ are scattering parameters describing the small signal interactions under spectral linearization around the LSOPs. $X_{p,[k_1,k_2,k_3];q,[k_1',k_2',k_3']}^{(S)}$ is a scattering parameter of type $S$ that accounts for the contribution to the frequency indexed as $[k_1,k_2,k_3]$ of the scattered wave at port $p$ from the $[k_1',k_2',k_3']$-th harmonic of the incident wave at port $q$. $X^{(T)}$ is a scattering parameter of type $T$. The definitions of the subscripts of $X^{(T)}$ are the same as those of the $X^{(S)}$ term except that they account for the contribution from harmonics of the conjugate of the incident wave. The existence of a scattering parameter of type $T$ is due to the nonanalyticity of the spectral mapping from the time domain to the frequency domain~\cite{ref23}. The sum runs over all $q$ and all integers $k_1',k_2',k_3'$.

\section{Volterra Kernel Extraction from X-parameters}\label{Sec4}
\subsection{Determination of Volterra Kernels}
Volterra kernels are transfer functions of a nonlinear system and are widely used to characterize a nonlinear system with memory. Previously, Volterra kernels are calculated by the harmonic input method \cite{ref22, ref22_1, ref25}. However, the determination process is very tedious and time-consuming. It requires the generation of input tones, waiting for steady-state, sampling the output, computation of Fourier transfer to get the output response at distinct frequencies.  When frequency sweep is needed, the determination task becomes extremely challenging especially for high-order kernels. On the other hand, X-parameters can be conveniently obtained from harmonic-balance (HB) simulation or measured by modern nonlinear vector network analyzers (NVNA). Inspired by the formalism of the incommensurate multi-tone X-parameters, we propose a systematic method to obtain the Volterra series representation of X-parameters based on the concept of harmonic input method. In this work, we take the negative frequencies into account by adding the complex conjugation terms and give the general representation of the output responses due to harmonic inputs. In general, to get the complete description of the $M$th-order Volterra kernel, an $M$-tone excitation is required. Suppose the input signal is the superposition of $M$ incommensurate tones
\newcounter{mytempeqncnt}
\setcounter{mytempeqncnt}{\value{equation}}
\setcounter{equation}{9}
\begin{eqnarray}\label{eq10}
u(t)=\sum_{m=1}^{M}\frac{V_m}{2}e^{j\omega_mt} + \text{c.c.}= \sum_{\substack{m=-M \\  m\neq 0}}^{M}\frac{V_m}{2}e^{j\omega_mt}
\end{eqnarray}
where ``c.c." denotes the complex conjugate terms, and $\omega_{-m}=-\omega_m$, and $V_{-m}=V_{m}^{*}$, $m$ is an integer. Substituting the input representation \eqref{eq10} into \eqref{eq2}, the $n$th-order output is
\begin{eqnarray}\label{eq11}
y_n(t)=\frac{1}{n!}\int_{-\infty}^{+\infty}\ldots\int_{-\infty}^{+\infty}h_n(\tau_1,\ldots,\tau_n) \cdot \qquad \qquad \quad \; \\ \nonumber
\qquad \prod_{i=1}^n \left[ \frac{V_1}{2}e^{j\omega_1 (t-\tau_i)}  + \ldots+ \frac{V_M}{2}e^{j\omega_M (t-\tau_i)} + \text{c.c.} \right]d\bar{\tau} \\ \nonumber =\frac{1}{n!}\sum_{m_1=-M}^{M}\ldots\sum_{m_n=-M}^{M}\left[\prod_{i=1}^n\frac{V_{m_i}}{2}\right]\cdot \qquad \qquad \quad \ \\ \nonumber
H_n^{sym}\left(\omega_{m_1},\ldots,\omega_{m_n}\right)
\cdot\exp{\left(j\sum_{i=1}^n\omega_{m_i}t\right)} \qquad \quad  \nonumber
\end{eqnarray}
where $m_i$ is an integer and $m_i\neq 0$, $d\bar{\tau}=d\tau_1\ldots d\tau_n$. $H_n^{sym}\left(\cdot\right)$ in \eqref{eq11} is defined as the $n$th symmetric frequency domain Volterra kernel or transfer function. The symmetric kernel satisfies
\begin{eqnarray}\label{eq12}
H_n^{sym}\left(\omega_1,\omega_2,\ldots,\omega_n\right)=H_n^{sym}\left(\omega_{m_1},\omega_{m_2},\ldots,\omega_{m_n}\right)
\end{eqnarray}
where the subscript of the argument $m_i$ denotes any permutation of the integers $1,\ldots, n$. It can be obtained by setting~\cite{ref25}
\begin{eqnarray}\label{eq13}
H_n^{sym}\left(\cdot\right)=\frac{1}{n!}\sum_{\substack{\textrm{all permutations of} \\ \{\omega_{m_1},\ldots,\omega_{m_n}\}}} H_n^{asym}\left(\omega_{m_1},\ldots,\omega_{m_n}\right)
\end{eqnarray}
In this paper, we use symmetric Volterra kernel. The superscript ``$sym$" is omitted for simplicity. Some of the values $\omega_{m_1},\ldots,\omega_{m_n}$ may be repeated. Thus, many terms in \eqref{eq11} contain identical exponents. Taking a three-tone excitation ($M=3$) as an example, the third-order output $y_3(t)$ in \eqref{eq11} contains $216$ third-order Volterra kernels. The arguments $\omega_{m_i}$ of these kernels $H_3\left(\omega_{m_1},\omega_{m_2},\omega_{m_3}\right)$ can be $\pm\omega_1$, $\pm\omega_2$ and $\pm\omega_3$. The Volterra kernels included in $y_3(t)$ are $H_3\left(-\omega_1,\omega_1,\omega_1\right)$, $H_3(\omega_1,-\omega_1,\omega_1)$, $H_3(\omega_1,\omega_1,-\omega_1)$, $H_3(\omega_1,\omega_2,-\omega_2)$, $H_3(\omega_1,-\omega_2,\omega_2)$, $H_3(\omega_2,\omega_1,-\omega_2)$, $H_3(-\omega_2,\omega_1,\omega_2)$, $H_3(\omega_2,-\omega_2,\omega_1)$, $H_3(-\omega_2,\omega_2,\omega_1)$, etc. Due to the permutation symmetry, the first three kernels are the same and corresponding to the output frequency $\omega=\omega_{m_1}+\omega_{m_2}+\omega_{m_3}=2\omega_1-\omega_1=\omega_1$. The next six kernels are the same and also corresponding to the output frequency $\omega=\omega_1+\omega_2-\omega_2=\omega_1$. These symmetric Volterra kernels coinciding at the same frequency can be collected together. We introduce a concise kernel  $G_{[k_1+r_1,r_1],\ldots,[k_M+r_M,r_M]}\left(\omega_1,\ldots,\omega_M\right)$ to denote all symmetric kernels at frequency $k_1\omega_1+k_2\omega_2+\dots+k_M\omega_M$ ($k_m$ is an integer and $0\le |k_m|\le n $).

$G_{[k_1+r_1,r_1],\ldots,[k_M+r_M,r_M]}\left(\omega_1,\ldots,\omega_M\right)$ is $H_n\big(\omega_{m_1},\ldots,$ $\omega_{m_n}\big)$ with
\begin{eqnarray}\label{eq14}
k_1+2r_1+\ldots+k_M+2r_M &=& n \\ \nonumber
\textrm{first } k_1+r_1 \textrm{ of } \omega_{m_i} &=& +\omega_1 \\ \nonumber
\textrm{next }  r_1 \textrm{ of } \omega_{m_i} &=& -\omega_1 \\ \nonumber
& \vdots & \\ \nonumber
\textrm{next } k_M+r_M \textrm{ of } \omega_{m_i} &=& +\omega_M \\ \nonumber
\textrm{next }  r_M \textrm{ of } \omega_{m_i} &=& -\omega_M  \nonumber
\end{eqnarray}
The number of arguments of $G(\cdot)$ is equal to the number of excitation tones. The number of arguments of $H(\cdot)$ is $k_1+2r_1+\ldots+k_M+2r_M$. The $m$th subscript of $G(\cdot)$ kernel $[k_m+r_m,r_m]$ corresponds to the arguments $\omega_m$ of $H(\cdot)$ (with $k_m+r_m$ positive $\omega_m$ and $r_m$ negative $\omega_m$ arguments). Hence, in the three-tone excitation example, the first three kernels, e.g., $H_3(\omega_1,\omega_1,-\omega_1)$,  can be rewritten as $G_{[1+1,1],[0+0,0],[0+0,0]}(\omega_1,\omega_2,\omega_3)$ (with $k_1=r_1=1$, $k_2=r_2=k_3=r_3=0$ in the general representation of $G(\cdot)$). That is, the first $k_1+r_1=2$ arguments of $H_3$ are $\omega_1$; the next $r_1=1$ argument is $-\omega_1$. Similarly, the next six kernels, e.g., $H_3(\omega_1,\omega_2,-\omega_2)$, can be denoted as $G_{[1+0,0],[0+1,1],[0+0,0]}(\omega_1,\omega_2,\omega_3)$ (with $k_1=r_2=1$, $r_1=k_2=k_3=r_3=0$ in the general representation of $G(\cdot)$). That is, the first $k_1+r_1=1$ argument is $\omega_1$; the next $r_1=0$ argument is $-\omega_1$; the next $k_2+r_2=1$ argument is $\omega_2$; and the next $r_2=1$ argument is $-\omega_2$. By using the new kernel representation and collecting all terms at frequency $k_1\omega_1+k_2\omega_2+\dots+k_M\omega_M$, we get a simplified representation of \eqref{eq11}
\begin{eqnarray}\nonumber \label{eq15_1}
y_n(t)=\!\sum_{r_1}\ldots\sum_{r_M}\Bigg[\left(\frac{V_1}{2}\right)^{k_1+r_1}\!\!\left(\frac{V_1^*}{2}\right)^{r_1}\!\!\ldots \left(\frac{V_M}{2}\right)^{k_M+r_M} \\ \nonumber
\left(\frac{V_M^*}{2}\right)^{r_M}\Bigg]\cdot\Bigg[(k_1+r_1)!r_1!\ldots(k_M+r_M)!r_M!\Bigg]^{-1} \quad \\ \nonumber
\cdot G_{[k_1+r_1,r_1],\ldots,[k_M+r_M,r_M]}\left(\omega_1,\ldots,\omega_M\right) \;\;  \\ \nonumber \cdot\exp{\left[j(k_1\omega_1+\ldots+k_M\omega_M)t\right]} \nonumber \quad \; \; \quad \textrm{(15)}
\end{eqnarray}
where $r_m$ are nonnegative integer indices that satisfy $k_1+2r_1+k_2+2r_2+\ldots+k_M+2r_M=n$ since $G_{[k_1+r_1,r_1],\ldots,[k_M+r_M,r_M]}(\cdot)$ is the $n$th-order kernel. The number of symmetric kernels denoted by $G_{[k_1+r_1,r_1],\ldots,[k_M+r_M,r_M]}(\cdot)$ is $\frac{n!}{(k_1+r_1)!r_1!\ldots(k_M+r_M)!r_M!}$. When $k_m<0$, the signs of $\omega_m$ are reversed, e.g., if $k_1<0$, the first $|k_1|+r_1$ arguments of $H_n$ are $-\omega_1$ and the next $r_1$ arguments are $+\omega_1$. For example, $H_3(-\omega_1,\omega_2,\omega_2)$ corresponds to the output frequency $\omega=-\omega_1+2\omega_2$. It is denoted by $G_{[-1+0,0],[2+0,0],[0+0,0]}$ with $k_1=-1<0$. Then we reverse the sign of $\omega_1$. Now the first $|k_1|+r_1=1$ argument becomes $-\omega_1$.

The general representation of the total output $y(t)$ due to an $M$-tone excitation is similar to \eqref{eq15_1} except that all $r_m$ go from zero to infinity.
\setcounter{mytempeqncnt}{\value{equation}}
\setcounter{equation}{15}
\begin{eqnarray}\nonumber\label{eq16}
y(t)=\!\sum_{r_1=0}^{\infty}\!\ldots\!\!\sum_{r_M=0}^{\infty}   \Bigg[\left(\frac{V_1}{2}\right)^{k_1+r_1}\!\!\left(\frac{V_1^*}{2}\right)^{r_1}\!\!\ldots \left(\frac{V_M}{2}\right)^{k_M+r_M} \\ \nonumber
\left(\frac{V_M^*}{2}\right)^{r_M}\Bigg]\cdot\Bigg[(k_1+r_1)!r_1!\ldots(k_M+r_M)!r_M!\Bigg]^{-1} \quad \\ \nonumber
\cdot G_{[k_1+r_1,r_1],\ldots,[k_M+r_M,r_M]}\left(\omega_1,\ldots,\omega_M\right)  \;\; \\ \nonumber \cdot\exp{\left[j(k_1\omega_1+\ldots+k_M\omega_M)t\right]} \nonumber \quad \; \; \quad \textrm{(16)}
\end{eqnarray}
The Volterra kernels can be determined once the magnitudes and phases of the corresponding frequency components of the output signal $Y(k_1\omega_1+\ldots+k_M\omega_M)$ are known. The information is provided by X-parameters. Once we know the mapping relationship between the output response due to an $m$-tone excitation and the $m$-tone X-parameters, we can calculate the $m$th-order Volterra kernels.

\subsection{Relationship between Volterra Kernels and X-parameters}
As discussed above, the determination of the $M$-th Volterra kernel requires an $M$-tone input signal. With multi-tone input, mixing products will occur. The maximum mixing order $M_0$ is defined as the maximum order of the intermodulation terms included in the output frequency list. For example, assume there are two fundamental tones $\omega_1$ and $\omega_2$. If $M_0=0$ or $1$, no mixing products are included in the output frequency list; if $M_0=2$, the $\pm|\omega_1\pm\omega_2|$ intermodulation terms are included; if $M_0=3$, additional $\pm|2\omega_1\pm\omega_2|$ and $\pm|\omega_1\pm 2\omega_2|$ terms are included. For better presentation, the $M=3$ case is used to illustrate the relationship between Volterra kernels and X-parameters. The maximum order of each tone is $M$ and the maximum mixing order $M_0$ is also set to $M_0=M$. In addition, thanks to the time invariant property~\cite{ref23}, the initial phase of each tone is set to zero ($V_m=V_m^*$). Based on \eqref{eq9} and \eqref{eq14}, we can link Volterra kernels with X-parameters by equating the same frequency component of the output signal, i.e., by setting $Y(k_1\omega_1+k_2\omega_2+k_3\omega_3)=B_{p,[k_1,k_2,k_3]}$. Table \ref{Tab1} gives the mapping between Volterra kernels and X-parameters of the first four frequency components.
\begin{table*}[htbp]
\caption{The Mapping between Volterra Kernels and X-Parameters of a Two Ports Network}
\label{Tab1}
\centering
\begin{tabular}{>{\centering}m{1.8cm}>{\centering}m{2.8cm} >{\centering}m{8cm} >{\centering\arraybackslash}m{1.8cm} }
\hline
Frequency\\index $[k_1,k_2,k_3]$ & Phasor \\ $Y(k_1\omega_1+k_2\omega_2+k_3\omega_3)$ & Volterra Kernels & X-parameters  \\ \hline
$[0,0,1]$ & $Y(\omega_3)$ & $\frac{V_3}{2}H_1(\omega_3)+\frac{V_3^3}{16}H_3(\omega_3,\omega_3,-\omega_3)+
\frac{V_3V_2^2}{8}H_3(\omega_3,\omega_2,-\omega_2)+\frac{V_3V_1^2}{8}H_3(\omega_3,\omega_1,-\omega_1)+
\frac{V_3^5}{384}H_5(\omega_3,\omega_3,\omega_3,-\omega_3,-\omega_3)+\dots$ & $B_{2,[0,0,1]}$ \\
$\quad$ & $\quad$ & $\quad$ & $\quad$ \\
$[0,0,2]$ & $Y(2\omega_3)$ & $\frac{V_3^2}{8}H_2(\omega_3,\omega_3)+\frac{V_3^4}{96}H_4(\omega_3,\omega_3,\omega_3,-\omega_3)+
\frac{V_3^2V_2^2}{32}H_4(\omega_3,\omega_3,\omega_2,-\omega_2)+\dots$ & $B_{2,[0,0,2]}$ \\
$\quad$ & $\quad$ & $\quad$ & $\quad$ \\
$[0,0,3]$ & $Y(3\omega_3)$ & $\frac{V_3^3}{48}H_3(\omega_3,\omega_3,\omega_3)+\frac{V_3^5}{768}H_5(\omega_3,\omega_3,\omega_3,\omega_3,-\omega_3)+
\frac{V_3^3V_2^2}{192}H_5(\omega_3,\omega_3,\omega_3,\omega_2,-\omega_2)+\dots$ & $B_{2,[0,0,3]}$ \\
$\quad$ & $\quad$ & $\quad$ & $\quad$ \\
$[0,1,-2]$ & $Y(\omega_2-2\omega_3)$ & $\frac{V_2V_3^2}{16}H_3(\omega_2,-\omega_3,-\omega_3)+\frac{V_2V_3^4}{192}H_5(\omega_2,-\omega_3,-\omega_3,-\omega_3,\omega_3)+\dots $ & $B_{2,[0,1,-2]}$ \\
\hline
\end{tabular}
\end{table*}
As shown in Table \ref{Tab1}, each frequency component contains the contribution from numerous Volterra kernels. Take the output frequency with index $[k_1,k_2,k_3]=[0,0,1]$ as an example. The output frequency is $\omega=k_1\omega_1+k_2\omega_2+k_3\omega_3=\omega_3$. According to \eqref{eq16}, the phasor of this frequency component can be written as:
\setcounter{mytempeqncnt}{\value{equation}}
\setcounter{equation}{16}
\begin{eqnarray}\nonumber \label{eq17}
Y(\omega_3)\!\!\!\!\! &=&\!\!\!\!\!\!\! \sum_{r_1=0}^{\infty}\sum_{r_2=0}^{\infty}\sum_{r_3=0}^{\infty}
\left[\frac{(V_1/2)^{2r_1}(V_2/2)^{2r_2}(V_3/2)^{1+2r_3}}{(r_1!)^2(r_2!)^2(1+r_3)!r_3!}\right] \\ \nonumber
&& \cdot G_{[0+r_1,r_1],[0+r_2,r_2],[1+r_3,r_3]}\left(\omega_1,\omega_2,\omega_3\right) \\ \nonumber
&=& B_{p,[0,0,1]} \qquad \; \qquad \qquad \qquad \qquad  \qquad \qquad \quad \textrm{(17)} \nonumber
\end{eqnarray}
when $r_1$, $r_2$ and $r_3$ run from zero to infinity, we can list the Volterra kernels included in \eqref{eq17} according to \eqref{eq14}
\setcounter{mytempeqncnt}{\value{equation}}
\setcounter{equation}{17}
\begin{eqnarray}\label{eq18}
r_1=r_2=r_3=0:  && H_1(\omega_3) \\ \nonumber
r_1=r_2=0 \textrm{ and } r_3=1:  && H_3(\omega_3,\omega_3,-\omega_3) \\ \nonumber
r_1=r_3=0 \textrm{ and } r_2=1:  && H_3(\omega_2,-\omega_2,\omega_3) \\ \nonumber
r_2=r_3=0 \textrm{ and } r_1=1:  && H_3(\omega_1,-\omega_1,\omega_3) \\ \nonumber
r_1=r_2=0 \textrm{ and } r_3=2:  && H_5(\omega_3,\omega_3,\omega_3,-\omega_3,-\omega_3) \\ \nonumber
\vdots \qquad \qquad  && \qquad \qquad  \vdots
\end{eqnarray}
Consequently, \eqref{eq17} can be expanded and rewritten as
\begin{eqnarray} \label{eq19}
Y(\omega_3)&=&\frac{V_3}{2}H_1(\omega_3)+\frac{V_3^3}{16}H_3(\omega_3,\omega_3,-\omega_3)+ \\ \nonumber
&&\frac{V_3V_2^2}{8}H_3(\omega_2,-\omega_2,\omega_3)+ \\ \nonumber
&&\frac{V_3V_1^2}{8}H_3(\omega_1,-\omega_1,\omega_3)+ \\ \nonumber
&&\frac{V_3^5}{384}H_5(\omega_3,\omega_3,\omega_3,-\omega_3,-\omega_3)+\ldots \\ \nonumber
&=&B_{p,[0,0,1]}
\end{eqnarray}
As presented in \eqref{eq19}, the output at frequency $\omega_3$ contains different orders of Volterra kernels: the linear term $H_1(\omega_3)$, the compression term $H_3(\omega_3,\omega_3,-\omega_3)\triangleq H_{3c}$, the desensitization terms $H_3(\omega_2,-\omega_2,\omega_3)\triangleq H_{3d2}, H_3(\omega_1,-\omega_1,\omega_3)\triangleq H_{3d1}$ and other higher-order terms with $n>3$. We need to separate these kernels.

\subsection{Separation of Volterra Kernels}
In general, the magnitude of high-order output decreases drastically as the magnitude of the input signal decreases even though the magnitude of higher-order kernel may be large. For instance, supposing the input signal is denoted as $\alpha u(t)$ (with $\alpha u(t)<1$), then the $n$th-order output $y_n(t)$ is proportional to $\alpha^n u^n(t)$ according to \eqref{eq2}. Thus, if the input is reduced by $3\,$dB, $y_1(t)$ falls by $3\,$dB, $y_2(t)$ falls by $6\,$dB and so on. The separation of Volterra kernels makes use of this property. Provided that the magnitude of the input signal $u(t)$ is smaller than some upper bound, the high order terms above $M$ are negligible. By setting an input signal with suitable magnitude, the infinite summation for each frequency component is truncated to a finite one by ignoring the higher-order terms.
Then the kernels can be separated based on the least square linear regression method \cite{ref26} in the frequency domain. The basic idea consists of  arranging the magnitudes of the input tones so that a matrix can be constructed for the kernels. Take \eqref{eq19} as an example and ignore the higher-orders with $n>3$ by changing $V_3$, a matrix equation is constructed:
\begin{eqnarray}\label{eq20}
\left[ \begin{array}{cccc}  \frac{V_3^{(1)}}{2}  & \frac{\left(V_3^{(1)}\right)^3}{16}  & \frac{V_3^{(1)}V_2^2}{8}  & \frac{V_3^{(1)}V_1^2}{8} \\
\frac{V_3^{(2)}}{2}  & \frac{\left(V_3^{(2)}\right)^3}{16}  & \frac{V_3^{(2)}V_2^2}{8}  & \frac{V_3^{(2)}V_1^2}{8} \\
\frac{V_3^{(3)}}{2}  & \frac{\left(V_3^{(3)}\right)^3}{16}  & \frac{V_3^{(3)}V_2^2}{8}  & \frac{V_3^{(3)}V_1^2}{8} \\
\frac{V_3^{(4)}}{2}  & \frac{\left(V_3^{(4)}\right)^3}{16}  & \frac{V_3^{(4)}V_2^2}{8}  & \frac{V_3^{(4)}V_1^2}{8} \\
\end{array} \right]
\left[ \begin{array}{c}  H_1 \\  H_{3c}  \\ H_{3d2}  \\  H_{3d1}  \\
\end{array} \right]  \\ \nonumber
= \left [\begin{array}{cccc} Y(\omega_3)^{(1)} & Y(\omega_3)^{(2)} & Y(\omega_3)^{(3)} & Y(\omega_3)^{(4)} \\
\end{array}\right ]^T
\end{eqnarray}
where $V_3^{(i)}$, $i=1,\ldots,4$, are some properly chosen magnitudes of the third tone and $Y(\omega_3)^{(i)}$ are the corresponding phasors of the frequency component $\omega_3$. Different Volterra kernels are the solutions of \eqref{eq20}. It is essential that the magnitudes of the input tones are properly chosen. They should be smaller than some upper bound  so that lower-order Volterra kernels will not be skewed by high order terms but not so small that the higher-order terms will be buried in the noise. One improvement is to add additional input magnitudes and use the least square solutions of the resulting overdetermined equation as the estimate of the kernels. The magnitudes of the other two tones $V_1$ and $V_2$ are also changed to form the overdetermined matrix equation. However, with the increasing number of higher order Volterra kernels being included, the kernel separation becomes much harder even with the least square method. The least square method may give poor estimates of the kernels. One remedy is that we can first get good estimates of low order kernels by properly choosing small magnitudes of input tones; then conduct an additional least square estimate for the higher order kernels by setting low order kernels as knowns to reduce the error propagation.

\subsection{Design of Input Signal}
To obtain a complete description of Volterra kenels $H_1(\omega_1)$, $H_2(\omega_1,\omega_2)$ and $H_3(\omega_1,\omega_2,\omega_3)$, frequency sweep along axes $\omega_1$, $\omega_2$ and $\omega_3$ in the interested region is required. In addition, for convenient kernel separation, output frequencies are required to be distinct from each other. Hence, careful attention must be paid in choosing the frequency components included in the input signal. With incommensurate input tones, this condition is satisfied automatically. However, it is difficult to ensure the incommensurate condition for all combinations of sweeping frequencies. In practice, the incommensurate requirement can be relaxed as follows: suppose the number of frequency sweeping points along the $\omega_{\alpha}$ axis is $N_\alpha$ with $\alpha=1,\,2,\,3$ ($M=3$ case); by carefully choosing the input frequencies $\omega_{1,i}$, $\omega_{2,j}$, and $\omega_{3,k}$, the output frequencies will not overlap with each other. Here, $i,\,j$ and $k$ are integers with $1\le i\le N_1$, $1\le j\le N_2$, and $1\le k\le N_3$. Hence, we need to choose the input frequencies such that
\begin{eqnarray}\label{eq21}
k_1\omega_{1,i}+k_2\omega_{2,j}+k_3\omega_{3,k}= k_1^\prime\omega_{1,i}+k_2^\prime\omega_{2,j}+k_3^\prime\omega_{3,k} \\ \nonumber
\textrm{ iff   } k_1=k_1^\prime,\,k_2=k_2^\prime \textrm{ and }k_3=k_3^\prime \nonumber
\end{eqnarray}
Equation \eqref{eq21} must be satisfied for all $k_1$, $k_2$, $k_3$, $k_1^\prime$, $k_2^\prime$, $k_3^\prime\in\{0,\pm1,\pm2,\ldots,\pm M_0\}$.

\subsection{Notation of Output Frequency}
For systems with real input and output signals in the time domain, the spectra in the frequency domain are double sided with conjugate symmetry, e.g., $H_3(\omega_1,\omega_2,-\omega_3)=H_3^*(-\omega_1,-\omega_2,\omega_3)$. Hence, the spectra contain redundant information by a factor of two. The output frequency is determined by indices $k_1,\,k_2,$ and $k_3$, and by values of input frequencies. For some set of $[k_1,k_2,k_3]$, the output frequency can be negative. The information about these negative frequencies can be obtained from the corresponding positive frequencies with indices $[-k_1,-k_2,-k_3]$. Here, we present an indexing scheme that includes all output frequencies and meanwhile removes the redundancy. Supposing the highest-order of Volterra kernels is $M=3$ and the maximum mixing order $M_0=M$, then the indices of the output frequency $[k_1,k_2,k_3]$ are arranged according to the following scheme:
\begin{enumerate}
\item the summation of absolute values of all indices is less than or equal to the maximum mixing order (e.g., $|k_1|+|k_2|+|k_3|\le M_0$);
\item the first index is always nonnegative (e.g., $k_1\ge 0$);
\item the first nonzero index begins with positive number (e.g., if $k_1=k_2=0$, $k_3>0$);
\item the second nonzero index begins with the available minimum integer (e.g., if $k_1\ne0$, $k_2$ begins with $-(M_0-|k_1|)$ ) and so does the next nonzero index (e.g., if $k_1\ne0$ and $k_2\ne0$, $k_3$ begins with $-(M_0-|k_1|-|k_2|)$ );
\end{enumerate}
These conditions are listed in descending order according to priority.  By using the above indexing scheme, instead of recording all combinations of $[k_1,k_2,k_3]$, only 31 frequencies need to be recorded for the $M=M_0=3$ case (see Appendix~\ref{Secappend} for more details). When the index $[k_1,k_2,k_3]$ results in a negative frequency, the complex conjugate of the corresponding phasor is considered and its contribution is attributed to the corresponding positive frequency.

In addition, Volterra kernels also have the permutation symmetry property, e.g., $H_3(\omega_1,\omega_2,\omega_3) =  H_3 (\omega_1,\omega_3,\omega_2)  =  H_3 (\omega_2,\omega_1,\omega_3)  =  H_3 (\omega_2,\omega_3,\omega_1)  =  H_3 (\omega_3,\omega_1,\omega_2)  =  H_3 (\omega_3,\omega_2,\omega_1)$. For each triplet $(\omega_{1,i}, \omega_{2,j}, \omega_{3,k})$, as shown in Fig. \ref{Fig1}, it will determine $216$ points in the $H_3(\omega_1,\omega_2,\omega_3)$ space. Meanwhile, $36$ and $6$ points will be determined in the $H_2(\omega_1,\omega_2)$ and $H_1(\omega_1)$ space, respectively. Due to the permutation and conjugate symmetry properties of kernels, only $28$ points need to be recorded for the third-order kernel, and $9$ and $3$ points for the second-order and first-order kernels, respectively. This leads to an efficient Volterra kernel determination process.

\begin{figure}[htbp]
\begin{center}
\noindent
  \includegraphics[width=2.5in]{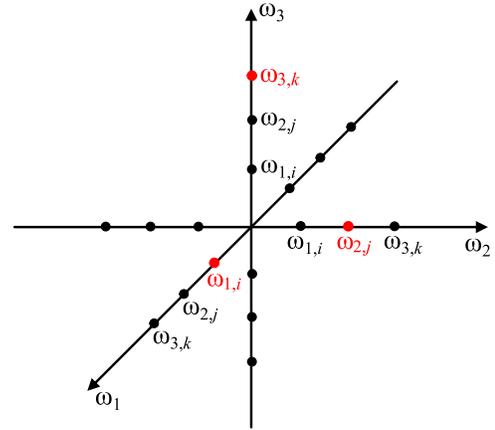}
  \caption{Distribution of points in the $H_3(\omega_1,\omega_2,\omega_3)$ space with one triplet $\left(\omega_{1,i},\omega_{2,j},\omega_{3,k}\right)$.}\label{Fig1}
\end{center}
\end{figure}


\section{Numerical Examples}\label{Sec5}
\subsection{A Benchmark Case}
\subsubsection{Description of the Nonlinear System}
The first numerical example is meanwhile to benchmark a nonlinear system with known Volterra kernels. Fig.~\ref{Fig2} shows the system diagram. $H_a$, $H_b$ and  $H_c$ are frequency domain transfer functions of linear-time-invariant systems. The symbol $\Pi$ denotes the time domain multiplication even though the subsystems are represented in the frequency domain. The V-I relationship at port $3$ of the multiplier is
\begin{eqnarray}\label{eq22}
i_3(t)=\left[v_1(t)v_2(t)-v_3(t)\right]/Z_f
\end{eqnarray}
where $v_j$ and $i_j$ are the voltage and current at port $j$ respectively, $j\in\{1,2,3\}$. $Z_f=R=50\,\Omega$.

\begin{figure}[htbp]
\begin{center}
\noindent
  \includegraphics[width=3.0in]{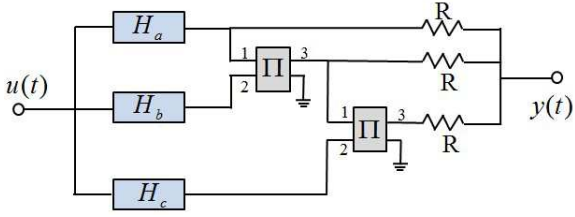}
  \caption{System diagram of a third-order nonlinear circuit.}\label{Fig2}
\end{center}
\end{figure}

Because there are two multipliers, the whole system has up to the third-order nonlinearity. In the frequency domain, Volterra kernels of the whole system have the following ideal analytical expression:
\begin{eqnarray}\label{eq23}
H_1(\omega_1)&=&c_1 H_a(\omega_1) \\ \nonumber
H_2(\omega_1,\omega_2)&=&c_2 \left[H_a(\omega_1)H_b(\omega_2)\right]_{\textrm{sym}} \\ \nonumber
H_3(\omega_1,\omega_2,\omega_3)&=&c_3 \left[H_a(\omega_1)H_b(\omega_2)H_c(\omega_3)\right]_{\textrm{sym}} \nonumber
\end{eqnarray}
where $[\cdot]_{\textrm{sym}}$ indicates a symmetrized Volterra kernel according to \eqref{eq13}; $c_i$ is a constant determined by circuit parameters (e.g., $Z_f$, $R$). However, due to the mismatch between different blocks, reflections will occur and slightly modify the ideal kernel representations. We set $H_a=H_b=H_c$ as the transfer function of a low pass filter (LPF) system. Fig.~\ref{Fig3} shows the magnitude of the transfer function $H_a$ which is equivalent to S-parameters $S_{21}$.
\begin{figure}[htbp]
\begin{center}
\noindent
  \includegraphics[width=2.5in]{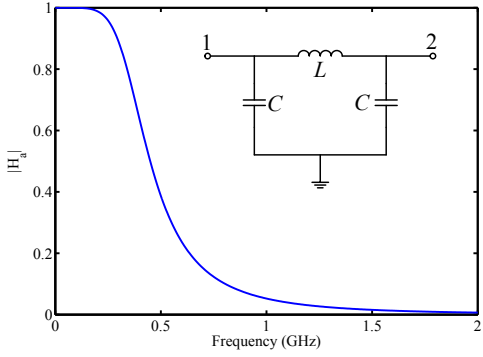}
  \caption{Magnitude of the transfer function $H_a$ of the low pass filter; the inset shows the circuit schematic diagram with $L=42.52\,\textrm{nH}$ and $C=8.5\,\textrm{pF}$.}\label{Fig3}
\end{center}
\end{figure}

\subsubsection{Frequency Domain Volterra Kernels}
Volterra kernels of the nonlinear system shown in Fig.~\ref{Fig2} are extracted from X-parameters.  The 3-tone X-parameters of the nonlinear system are generated by applying the ADS X-parameters generator~\cite{ref27}. The frequency sweep scheme of each axis is given in Table~\ref{Tab2}. Although the frequency step of each axis is $120\,$MHz, the equivalent frequency step is $40\,$MHz thanks to the symmetry properties of kernels.
\begin{table}[htbp]
\caption{Frequency Sweep Scheme for the 3-tone X-parameters Generation}
\label{Tab2}
\centering
\begin{tabular}{c c c c}
  \hline
  Frequency & Start & Step & Stop\\
  \hline
  $\omega_1$ & $7\,$MHz &  $120\,$MHz & $2.047\,$GHz \\
  $\omega_2$ & $41\,$MHz & $120\,$MHz & $2.081\,$GHz \\
  $\omega_3$ & $87\,$MHz & $120\,$MHz & $2.127\,$GHz \\
  \hline
\end{tabular}
\end{table}
To separate different order of Volterra kernels, the power of each input tone is set to $P_{in}=\{5,10\}\,\textrm{dBm}$. The proposed method described in Sec.~\ref{Sec4} is used to extract Volterra kernels $H_1$, $H_2$ and $H_3$. Fig.~\ref{Fig4} shows the magnitude and phase of the first-order Volterra kernel. They agree well with the S-parameters of the system with small signal input. Fig.~\ref{Fig5} and Fig.~\ref{Fig6} present the magnitudes and phases of the second-order Volterra kernel and a slice of the third-order Volterra kernel, respectively, with $\omega_3$ fixed at $0.5\,\textrm{GHz}$. They have reasonable distributions compared to the ideal analytical expression in \eqref{eq19}. In addition, they also agree with the permutation and conjugate symmetric properties. For example, the magnitude of $H_2$ has two symmetry planes: $\omega_1=\pm\omega_2$; the phase of $H_2$ has a symmetry plane $\omega_1=\omega_2$ and an anti-symmetry plane $\omega_1=-\omega_2$; while the third-order kernel loses the conjugate symmetry property since $\omega_3$ is fixed. It only has one symmetry plane $\omega_1=\omega_2$.

\begin{figure}[htbp]
\begin{center}
\noindent
  \includegraphics[width=3.5in]{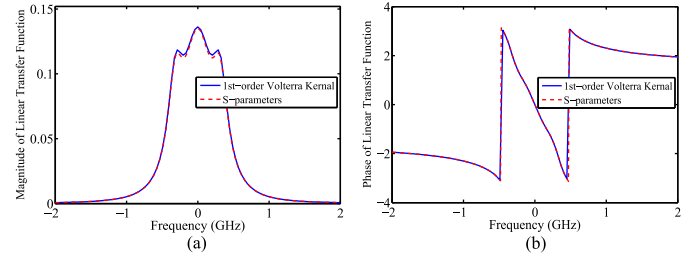}
  \caption{Magnitude and phase of the linear transfer function $H_1(\omega_1)$. (a) magnitude; (b) phase.}\label{Fig4}
\end{center}
\end{figure}
\begin{figure}[htbp]
\begin{center}
\noindent
  \includegraphics[width=3.5in]{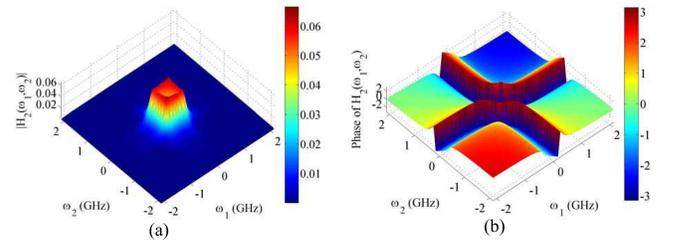}
  \caption{Magnitude and phase of the second-order Volterra kernel $H_2(\omega_1,\omega_2)$; (a) magnitude; (b) phase.}\label{Fig5}
\end{center}
\end{figure}
\begin{figure}[htbp]
\begin{center}
\noindent
  \includegraphics[width=3.5in]{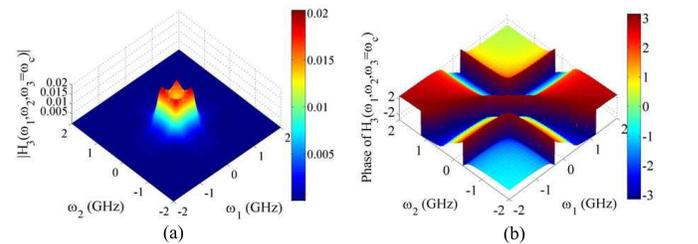}
  \caption{Magnitude and phase of the third-order Volterra kernel $H_3(\omega_1,\omega_2, $ $\omega_3 = \omega_c)$ with $\omega_c$ fixed to $0.5\,\textrm{GHz}$; (a) magnitude; (b) phase.}\label{Fig6}
\end{center}
\end{figure}

\subsubsection{Time Domain Output}
To validate the accuracy of the extracted Volterra kernels and to demonstrate the capability of Volterra series for time domain simulation with arbitrary input, the time domain outputs are presented. Without losing generality, the input is chosen to be a rectangular pulse as shown in the inset of Fig.~\ref{Fig7}(b). The magnitude is $V_0=1\,\textrm{V}$. The input has rich frequency components as shown in Fig.~\ref{Fig7}(a). The X-parameter macro-model requires numerous fundamental tones to represent the input~\cite{ref28}. In addition, once the input signal changes, one needs to regenerate the X-parameter macro-model for the same system since X-parameter is defined under certain LSOPs (fixed magnitude and phase of each fundamental tone). In contrast, the same Volterra series model can be used for arbitrary input. The linear, second-order and third-order time domain responses $y_1(t)$, $y_2(t)$ and $y_3(t)$  are given in Fig.~\ref{Fig7}(b), (c) and (d), respectively. They are calculated according to \eqref{eq5} and \eqref{eq7} with extracted Volterra kernels. The total response is the summation of both linear and nonlinear responses as indicated by \eqref{eq1}. Fig.~\ref{Fig8} presents the total response calculated by the Volterra series representation. It agrees very well with that obtained by ADS transient simulator with the original circuit model. However, as shown in Fig. 8, the linear response calculated by S-parameter does not agree with the response of the original circuit model. The higher-order Volterra kernels are crucial to capturing the nonlinearities of the systems. The linear response ignores the contributions of higher order Volterra kernels and hence it is incorrect.

\begin{figure}[htbp]
\begin{center}
\noindent
  \includegraphics[width=3.0in]{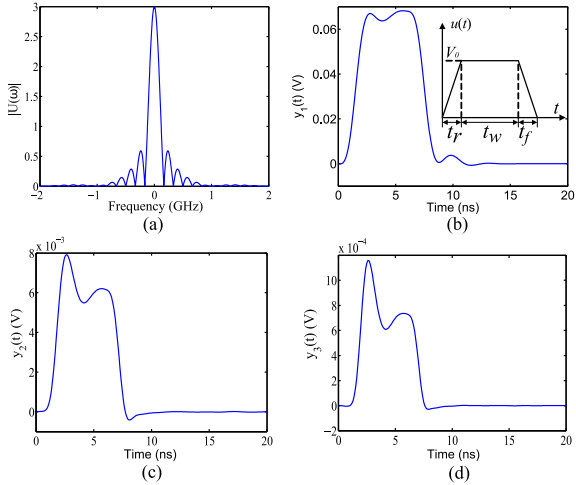}
  \caption{Input and output signal of the benchmark nonlinear system as shown in Fig.~\ref{Fig2}. (a) Spectrum of the rectangular pulse input.  Time domain output: (b) linear response $y_1(t)$; (c) second-order response $y_2(t)$; (d) third-order response $y_3(t)$. The inset in (b) shows the shape of the rectangular pulse with raise and fall time $t_r=t_f=1\,\textrm{ns}$, and width $t_w=5\,\textrm{ns}$.}\label{Fig7}
\end{center}
\end{figure}

\begin{figure}[htbp]
\begin{center}
\noindent
  \includegraphics[width=2.9in]{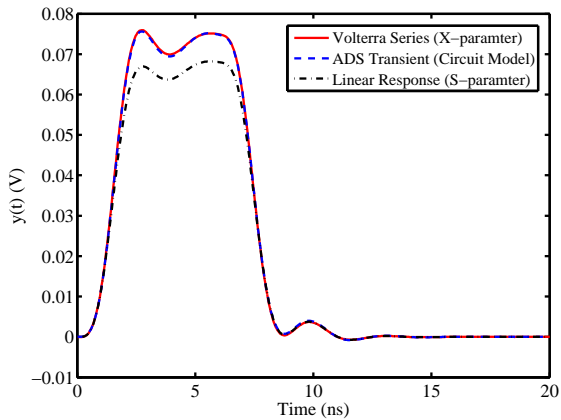}
  \caption{Comparison of the time domain output of the benchmark nonlinear system as shown in Fig.~\ref{Fig2} calculated by the Volterra series representation and ADS transient simulator.}\label{Fig8}
\end{center}
\end{figure}

\subsection{Low Noise Amplifier}
\subsubsection{Frequency Domain Volterra Kernel}
The second numerical example is a low noise amplifier (LNA). The amplifier schematic used in this work is taken from \emph{X-parameter generation tutorial} of the example directory of ADS \cite{ref27}. The inset in Fig.~\ref{Fig9}(a) shows the macro-model of the LNA. The saturation limit of input is $70\,\textrm{mV}$. The 3-tone X-parameters are generated with the same frequency sweep scheme shown in Tab.~\ref{Tab2}. The power of each tone is set to $P_{in}=\{-30,-20\}\,$dBm. The proposed method is used to extract Volterra kernels from X-parameters. Fig.~\ref{Fig9}(a) and (b) show the magnitude and phase of the first-order Volterra kernel $H_1$. They agree well with the small signal S-parameters. This is reasonable since X-parameters becomes S-parameters in the small signal limit. Fig.~\ref{Fig9}(c) and (d) present the magnitudes of the second-order Volterra kernel and a slice of the third-order Volterra kernel with $\omega_3=0.5\,\textrm{GHz}$, respectively. Again, $H_2$ has two symmetry planes due to permutation and conjugate symmetry while $H_3$ only has one symmetry plane due to the lost of the conjugate symmetry.
\begin{figure}[htbp]
\begin{center}
\noindent
  \includegraphics[width=3.4in]{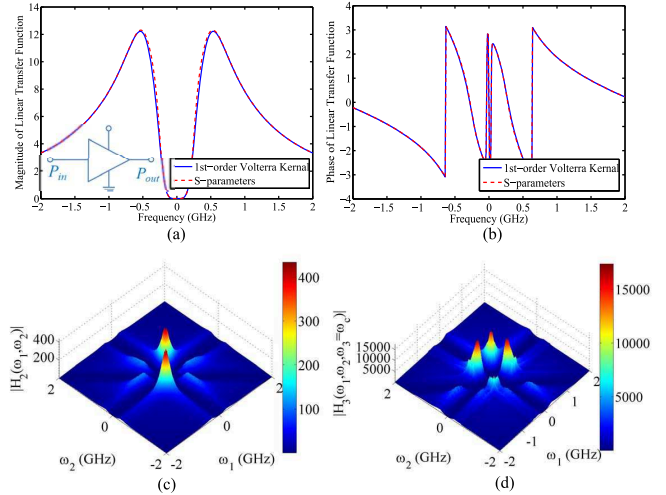}
  \caption{Volterra kernels of the low noise amplifier (LNA). (a) and (b) are the magnitude and phase of the linear transfer function (first-order Volterra Kernel $H_1(\omega_1)$); (c) magnitude of the second-order Volterra kernel $H_2(\omega_1,\omega_2)$; (d) a slice of the magnitude of the third-order Volterra kernel $H_3(\omega_1,\omega_2,\omega_3=\omega_c)$ with $\omega_c=0.5\,\textrm{GHz}$. The inset in (a) is the macro-model of the LNA.}\label{Fig9}
\end{center}
\end{figure}

\subsubsection{Time Domain Output}
After obtaining the Volterra kernels, different orders of time domain responses are calculated according to \eqref{eq5} and \eqref{eq7}. The rectangular pulse shown in the inset of Fig.~\ref{Fig7}(b) is used with $V_0=0.2\,\textrm{V}$. Fig.~\ref{Fig10} displays the total output $y(t)$ calculated based on the Volterra series representation. It captures the distortion due to nonlinearities and agrees well with the result simulated by the ADS transient simulator with the original circuit model.
\begin{figure}[htbp]
\begin{center}
\noindent
  \includegraphics[width=2.9in]{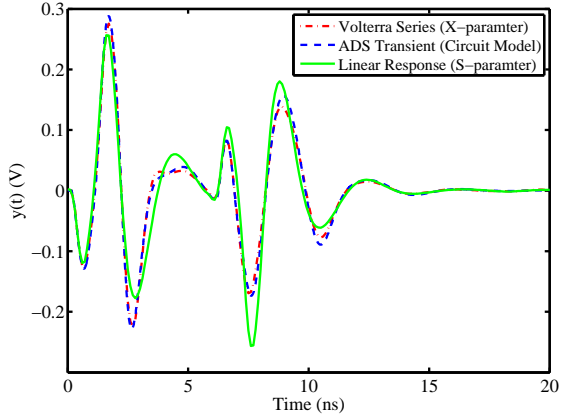}
  \caption{Comparison of the time domain output of the low noise amplifier calculated by the Volterra series representation and ADS transient simulator.}\label{Fig10}
\end{center}
\end{figure}


%
%

\section{Conclusion}\label{SecCon}
This paper presents a systematic method for extracting Volterra series representation from X-parameters. By completely separating different order of Volterra kernels based on the least square linear regression method, the complete description of Volterra kernels can be determined very efficiently. Time domain output can be obtained conveniently based on the determined Volterra kernels with arbitrary input. Numerical results show the capability of Volterra series representation for describing nonlinear devices in a broad input power region. The procedure for extracting Volterra kernels with the truncation order $M=3$ is illustrated in detail. The paper gives the general relationship between Volterra series and X-parameters and the method can be applied to the extraction of higher-order Volterra kernels.

\appendices
\section{Output Frequencies and Volterra Kernels for the Three-Tone Input Case}\label{Secappend}
For the three-tone input case, the output frequencies are the mixing of the input frequencies $\omega=k_1\omega_1+k_2\omega_2+k_3\omega_3$ and can be indexed as $[k_1,k_2,k_3]$. Suppose the maximum mixing order is $M_0=3$, the first-order output contains $3$ output frequencies and the corresponding $3$ Volterra kernels as listed in Tab.~\ref{Tab_app1}.

\begin{table}[htbp]
\renewcommand{\arraystretch}{1.3}
\caption{Frequencies and Volterra Kernels Included in the First-Order Output}
\label{Tab_app1}
\centering
\begin{tabular}{p{1.9cm}<{\centering}p{1.8cm}<{\centering}p{2.8cm}<{\centering}}
\hline
Frequency & Notation & Volterra Kernel \\ \hline
$\omega_1$ & [1,0,0] & $H_1(\omega_1)$ \\
$\omega_2$ & [0,1,0] & $H_1(\omega_2)$ \\
$\omega_3$ & [0,0,1] & $H_1(\omega_3)$ \\ \hline
\end{tabular}
\end{table}

The second-order output contains $9$ output frequencies and the corresponding $9$ Volterra kernels as shown in Tab.~\ref{Tab_app2}.

\begin{table}[htbp]
\renewcommand{\arraystretch}{1.3}
\caption{Frequencies and Volterra Kernels Included in the Second-Order Output}
\label{Tab_app2}
\centering
\begin{tabular}{p{1.9cm}<{\centering}p{1.8cm}<{\centering}p{2.8cm}<{\centering}}
\hline
Frequency & Notation & Volterra Kernel \\ \hline
$2\omega_1$ & [2,0,0] & $H_2(\omega_1,\omega_1)$ \\
$2\omega_2$ & [0,2,0] & $H_2(\omega_2,\omega_2)$ \\
$2\omega_3$ & [0,0,2] & $H_2(\omega_3,\omega_3)$ \\
$|\omega_1,\pm\omega_2|$ & $[1,\pm 1,0]$ & $H_2(\omega_1,\pm\omega_2)$ \\
$|\omega_2,\pm\omega_3|$ & $[0,1,\pm 1]$ & $H_2(\omega_2,\pm\omega_3)$ \\
$|\omega_1,\pm\omega_3|$ & $[1,0,\pm 1]$ & $H_2(\omega_1,\pm\omega_3)$ \\ \hline
\end{tabular}
\end{table}

Table~\ref{Tab_app3} shows the $22$ output frequencies and the corresponding $28$ Volterra kernels included in the third-order output.

\begin{table}[htbp]
\renewcommand{\arraystretch}{1.5}
\caption{Frequencies and Volterra Kernels Included in the Third-Order Output}
\label{Tab_app3}
\centering
\begin{tabular}{p{1.9cm}<{\centering}p{1.8cm}<{\centering}p{2.8cm}<{\centering}}
\hline
Frequency & Notation & Volterra Kernel \\ \hline
$\omega_1$ & [1,0,0] & $H_3(\omega_1,\omega_1,-\omega_1)$, $H_3(\omega_1,\omega_2,-\omega_2)$, $H_3(\omega_1,\omega_3,-\omega_3)$\\
$\omega_2$ & [0,1,0] & $H_3(\omega_2,\omega_1,-\omega_1)$, $H_3(\omega_2,\omega_2,-\omega_2)$, $H_3(\omega_2,\omega_3,-\omega_3)$ \\
$\omega_3$ & [0,0,1] & $H_3(\omega_3,\omega_1,-\omega_1)$, $H_3(\omega_3,\omega_2,-\omega_2)$, $H_3(\omega_3,\omega_3,-\omega_3)$ \\
$3\omega_1$ & [3,0,0] & $H_3(\omega_1,\omega_1,\omega_1)$ \\
$3\omega_2$ & [0,3,0] & $H_3(\omega_2,\omega_2,\omega_2)$ \\
$3\omega_3$ & [0,0,3] & $H_3(\omega_3,\omega_3,\omega_3)$ \\
$|\omega_1,\pm 2\omega_2|$ & $[1,\pm 2,0]$ & $H_3(\omega_1,\omega_2,\omega_2)$, $H_3(\omega_1,-\omega_2,-\omega_2)$ \\
$|2\omega_1,\pm \omega_2|$ & $[2,\pm 1,0]$ & $H_3(\omega_1,\omega_1,\omega_2)$, $H_3(\omega_1,\omega_1,-\omega_2)$  \\
$|\omega_2,\pm 2\omega_3|$ & $[0,1,\pm 2]$ & $H_3(\omega_2,\omega_3,\omega_3)$, $H_3(\omega_2,-\omega_3,-\omega_3)$ \\
$|2\omega_2,\pm \omega_3|$ & $[0,2,\pm 1]$ & $H_3(\omega_2,\omega_2,\omega_3)$, $H_3(\omega_2,\omega_2,-\omega_3)$ \\
$|\omega_1,\pm 2\omega_3|$ & $[1,0,\pm 2]$ & $H_3(\omega_1,\omega_3,\omega_3)$, $H_3(\omega_1,-\omega_3,-\omega_3)$ \\
$|2\omega_1,\pm \omega_3|$ & $[2,0,\pm 1]$ & $H_3(\omega_1,\omega_1,\omega_3)$, $H_3(\omega_1,\omega_1,-\omega_3)$ \\
$|\omega_1,\pm \omega_2\pm \omega_3|$ & $[1,\pm 1,\pm 1]$ & $H_3(\omega_1,\omega_2,\omega_3)$, $H_3(\omega_1,\omega_2,-\omega_3)$, $H_3(\omega_1,-\omega_2,\omega_3)$, $H_3(\omega_1,-\omega_2,-\omega_3)$ \\ \hline
\end{tabular}
\end{table}

It should be noticed that both the first-order and the third-order outputs contain the output frequencies $\omega_1$, $\omega_2$, $\omega_3$. Hence, the total number of the output frequencies is $31$.

\section*{Acknowledgment}
This work was supported in part by NSFC 61271158, US AOARD 124082 contracted through UTAR, Hong Kong UGC AoE/P-04/08, and by the US National Science Foundation Award 1218552. The authors thank Dr. Ngai Wong for the constructive suggestions and Keysight Technologies Inc., for supporting this work and providing the ADS X-parameter generation platform; especially, we thank David Root, Loren Betts, Steve Fulwider and Bill Wallace of Keysight for fruitful discussions, insightful comments and helpful suggestions. X-parameters is a registered trademark of Agilent Technologies.

\ifCLASSOPTIONcaptionsoff
  \newpage
\fi

\end{document}